\documentclass[aps,prl,twocolumn,showpacs,preprintnumbers,
amsmath,amssymb,superscriptaddress,showkeys]{revtex4}

\usepackage{graphicx} \usepackage{color} \usepackage{dcolumn}
\usepackage{bm}

\definecolor{violet}{rgb}{0.5,0,0.5}
\makeatletter \def\@dotsep{5} \makeatother

\newcommand{\tdp}{{\hbox{$\pmb{\pmb{\boldsymbol{\blacktriangledown}}}
 \mkern-20mu$\raise0.15ex\hbox{\textbf{- -}}}}}

\newcommand{\degree}{\ensuremath{^\circ}}

\begin{document}

\title{A novel technique to make Ohmic contact to a buried two-dimensional electron gas in a molecular-beam-epitaxy grown $GaAs/Al_{0.3}Ga_{0.7}As$ heterostructure with Mn $\delta$-doping}

\author{A. Bove}\affiliation{Physics Department, Duke University, Durham, NC 27708} \affiliation{Physics Department, Purdue University, West Lafayette, IN 47906 } 

\author{F. Altomare} \affiliation{NIST, 325 Broadway, Boulder, Colorado 80305} 

\author{N. B. Kundtz} \affiliation{Physics Department, Duke University, Durham, NC 27708}

\author{Albert M. Chang} \email{yingshe@phy.duke.edu} \affiliation{Physics Department, Duke University, Durham, NC 27708}

\author{Y. J. Cho} \author{X. Liu} \author{J. Furdyna} \affiliation{Department of Physics, University of Notre Dame, Notre Dame, Indiana 46556} 

\date{\today}

\begin{abstract}
We report on the growth and characterization of a new Diluted Magnetic Semiconductor (DMS) heterostructure that presents a Two-Dimensional Electron Gas (2DEG) with a Quantum Well (QW) carrier's density in the range $0.5\leq n \leq 1\times10^{12} cm^{-2}$ and a mobility $350\leq\mu\leq600 cm^{2}/Vs$ at $T\sim4.2K$. As far as we know this is the highest mobility value reported in the literature for GaMnAs systems. A novel technique was developed to make Ohmic contact to the buried 2DEG without destroying the magnetic properties of our crystal.
\end{abstract}


\maketitle

\section{Introduction}
The growth of high quality crystals (i.e. low impurity content) is one of the major challenges in the realization of GaMnAs-based ferromagnetic devices. One of the major lines of research in the fabrication of ferromagnetic devices is directed at achieving a high Curie Temperature $T_{C}$ in order to exploit the device characteristics at room temperature (\cite{2DHG}, \cite{2DHG2}). The ferromagnetic properties of GaMnAs depend on the carrier's density and Mn concentration \cite{DietlZener}. Therefore, in order to reach high critical temperatures $T_{C}$, it is necessary to fabricate devices with high carrier concentration and high Mn content. This is achieved by using Low Temperature MBE (LT-MBE) growth techniques that are capable of incorporating a high concentration of Mn in the GaMnAs layer while keeping the substrate temperature at around 300\degree C \cite{Ohno1992}. However, this technique has the drawback of producing devices with poor mobility due to the presence of many lattice defects in the low temperature grown layers and magnetic impurities all acting as scattering centers thus, heavily affecting the carrier's mean free path. In contrast with this line of research, our group has decided to pursue the fabrication of devices with low carrier concentration and high mobility with the ultimate objective of building the first ferromagnetic quantum dot and thus, studying the basic mechanism behind such ferromagnetic behavior. While these devices will only work at cryogenic temperatures, they could provide a proof of concept for the fabrication of devices with higher $T_{C}$.
In this article, we present a low carrier concentration ($0.5\leq n \leq 1\times10^{12} cm^{-2}$) GaMnAs QW contained in a $GaAs/Al_{0.3}Ga_{0.7}As$ heterostructure, specially designed to minimize the impurity content and maximize the carrier's mobility. We also describe a novel technique, which we developed, to make Ohmic contact to the buried 2DEG that preserves the ferromagnetic properties of the new heterostructure. Magneto-transport measurements show that the n-type carrier's mobility measured in the QW is in the range $350\leq\mu\leq600 cm^{2}/Vs$, which is the highest value, as far as we know, reported in the literature for GaMnAs systems. This heterostructure shows ferromagnetism when it is cooled at a temperature below $1.7K$ and when the quantum well sitting next to the Mn impurities is filled with electrons that are otherwise sitting at a heterojunction far below.

\section{Crystal Structure and Ohmic contact}

The MBE growth sequence from the bottom to the top is: a semi-insulating (SI) GaAs (001) substrate (see table 1 for the growth structure),  a 200 nm GaAs Si-doped layer, a 100 nm AlGaAs Si/Be-doped, a 10 nm SI-GaAs. All of these layers were grown while keeping the substrate temperature at 600\degree C to minimize the presence of impurities. The substrate temperature was, then, lowered to $T_{S}\sim$250\degree C, a 2/3 monolayer (ML) of Mn was grown, then a 10 nm AlGaAs buffer layer, then a 15 nm AlGaAs Be-doped layer, and finally a 5 nm SI-GaAs capping layer. The presence of the 10 nm Be-doped AlGaAs layer has two purposes: 1) it should reduce the presence of interstitial Mn which acts as a double donor and tends to couple anti-ferromagnetically with the substitutional Mn \cite{JungwirthRMP}, and 2) it limits the number of surface states that can act as ionized traps and thus, significantly affect the carrier's density and mobility. 
If the 2/3 ML of Mn atoms replace exactly 2/3 of a ML of Ga this would give us a concentration per unit cell of roughly $x\sim66.6\%$. From literature and from private conversations with Dr. Furdyna's group we can assume that the Mn layer spreads over 2 to 3 ML, thus, the nominal Mn concentration results to be $x\sim33.3\%$.
\begin{table}[t]
	\centering \caption{Heterostructure}
	\begin{tabular}[t]{ll}
						GaAs (001)& Substrate\\
						GaAs:Si& 100 nm\\
            AlGaAs:Be& 200 nm\\
					  SI-GaAs& 10 nm\\
					  LT-Mn& $\frac{2}{3}$ ML\\
					  LT-AlGaAs& 10 nm\\
					  LT-AlGaAs:Be& 15 nm\\
					  LT-GaAs& 5 nm\\
	\end{tabular}
	\label{tab:magneticstructure}
\end{table}
In this structure, a layer of 2D electron gas (2DEG) is always present at the interface of the Si-doped 200nm GaAs and the 100 nm AlGaAs:Si/Be layers, approximately 140 nm below the surface, while the 10 nm GaMnAs QW can be filled or emptied by gating.

The first major task is to make ohmic contact to the buried 2DEG without destroying the ferromagnetic properties. While it has been reported that moderate temperatures ($\sim$300\degree C) (See \cite{2DHG} and \cite{WangJAP2004}) for extended periods of time, lead to an increased $T_{C}$ and do not destroy the ferromagnetic properties of these structures, these temperatures are not sufficient to produce good Ohmic contacts. Instead, it is necessary to go at much higher temperatures, T$\sim$400\degree C. However, at these temperatures, the magnetism is destroyed. To preserve the ferromagnetic properties and at the same time make good ohmic contact to the buried 2DEG, we developed a novel technique which employs a thermal gradient along the sample. This method of annealing destroys the magnetic properties in the area of the contacts but preserves them in the center of the sample. To test our technique, ferromagnetic epilayers are very useful because they allow us to experiment with different annealing conditions in order to characterize the effect of the temperature on the ferromagnetic properties of the samples. Not only these properties can be easily monitored using the Giant Planar Hall Effect (GPHE) \cite{GPHE}, but in turn we can use the GPHE to establish when the ferromagnetic properties are lost. Epilayers show hole-mediated ferromagnetism, have a Curie Temperature $T_{C}\sim$30 K and since they are grown directly on the surface of the wafer they require no annealing to make Ohmic contact. An InZn alloy ($\sim95\%$ In and $\sim5\%$ Zn) is used to make electrical contact. 
\begin{figure}[ht]

\centering \includegraphics[clip, width=3.1in]{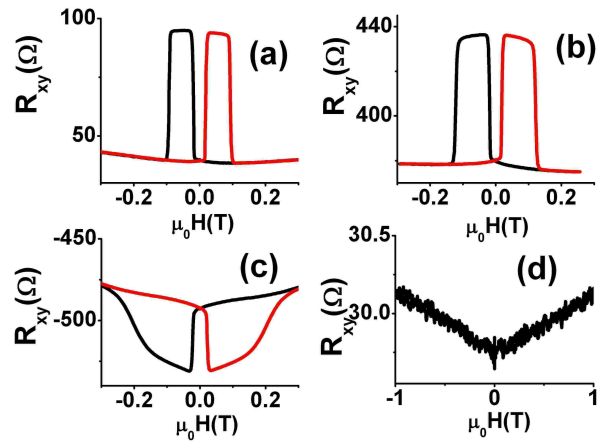}
\centering \caption{GPHE (Giant Planar Hall Effect \cite{GPHE}) at different annealing conditions. (a) No annealing. (b) Annealing at $T_{a}\sim320\degree C$ for 30 s. (c) Annealing at $T_{a}\sim370\degree C$ for 30 s (d) Annealing at $T_{a}\sim420\degree C$ for 30 s.}

\label{fig:GHPE}

\end{figure}
We defined a Hall bar geometry with etching and monitored the GPHE as a function of the annealing temperature (see figure 1) \cite{GPHE}. Figure 1(a) displays the GPHE measured on such a device. We then annealed the sample at different temperatures and measured the GPHE as a function of the annealing temperature. By increasing the annealing temperature, we observe an increase in the size of the hysteresis loop when $T_{a}\sim$320\degree C (figure 1(b)) and, for $T_{a}\sim$370\degree C, a rotation of the easy magnetization axis is observed (figures 1(c)). At the highest temperature used ($T_{a}\sim$420\degree C), the GPHE has completely disappeared (figures 1(d)). This behavior clearly demonstrated that this gradient annealing technique could work in principle. 

The next step was to reproduce the same results with a sample on which a thermal gradient had been created along its length. We took a new piece, $\sim$8 mm long and $\sim$2.5 mm wide, of the same wafer and deposited equally distanced InZn contacts starting from one of the short edges and ending on the other. 
\begin{figure}[ht]

\centering \includegraphics[clip, width=3.1in]{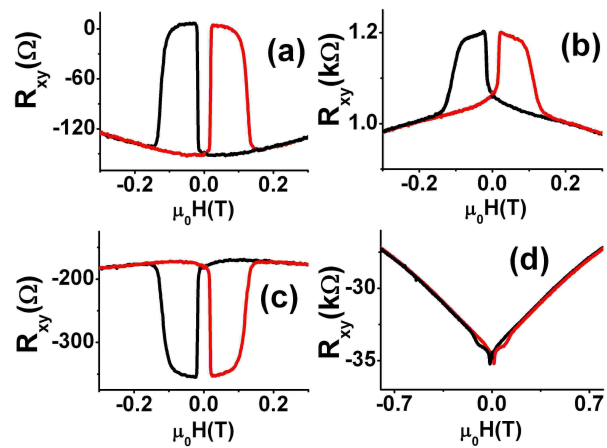}
\centering \caption{Only in the region closest to the heat source (displayed in d), the temperature increase destroys the ferromagnetism. The data displayed in (a)-(d) closely resemble those in figures 1(a-d) thus, demonstrating that the annealing process, in region far from the heat source, preserve the ferromagnetic properties of the sample.}

\label{fig:GHPE1}

\end{figure}
These contacts allowed us to monitor the GPHE as a function of the distance from the hottest point to the coolest, thus spatially probing the effect of increasing temperatures. We then placed our sample in a vertical position and on a hot plate such that one edge would be in thermal contact while the other would be kept at a much lower temperature by the nitrogen flow. The results are shown in figures 2(a-d). While in the central region of the sample, the GPHE is visible (figure 2(a)), moving towards the edge of the sample kept in thermal contact with the hot plate, (figures 2(b-c)) the shape of the hysteresis changes in a way that closely resembles the curves displayed in figures 1(a-d). This comparison allows us to put an upper bound on the temperature in the center of the sample to roughly 270\degree C. These results clearly prove that our strategy is effective. This is a very important result in itself because our methodology provides a way to make contact to a buried 2D gas without destroying the ferromagnetic properties of the sample and can, in principle, be used in other situations where a similar control of the temperature is necessary.

\section{Magneto-Transport Measurements}
To characterize our structure and to determine if our annealing technique could work on a real sample, we made ohmic contact, using pure In, to a cleaved piece of our crystal. We, then, lithographically defined a Hall bar with a 150 $\mu$m wide channel in the center of the sample. The voltage probes were narrow, 10 $\mu$m in width, to minimize their perturbation on the current flow. Magneto-transport measurements were carried out in the standard Hall-effect configuration using a lock-in amplifier at 3.7 Hz and 10 nA excitation current. 

At 4.2K, the sign of the Hall coefficient in figure 3(c) and the gate voltage dependence of the carrier density in figure 3(d) clearly indicated that our structure contains n-type carriers. This was confirmed by comparing the hall coefficient's sign of this sample with the one of a sample with a known carrier type. This sample showed a carrier's density $n\sim1.08\times10^{12}cm^{-2}$ and, with a resistivity per square of $R_{\square}\sim$10k$\Omega$, had a mobility $\mu\sim575cm^{2}/Vs$ (The density and mobility's values are associated with the carriers residing in the HJ as it is shown in the last section of this letter). It should also be noticed that imperfections in the geometry slightly affects the anti-symmetric behavior of the $R_{xy}$ at 4.2 K (above $T_{C}$), as it can be seen in figure 3(a). This effect is accentuated when the sample is cooled below the Curie temperature. To determine the carrier's density we, first, removed the magnetic field's lag by subtracting the exponential time dependence of the lag, and, then, we anti-symmetrized to remove the symmetric component associated with the imperfections in the geometry. Figures 3(a-b) show no sign of ferro-magnetism at 4.2 K either when an in-plane or perpendicular magnetic field was applied.
\begin{figure}[ht]

\centering \includegraphics[clip, width=3.1in]{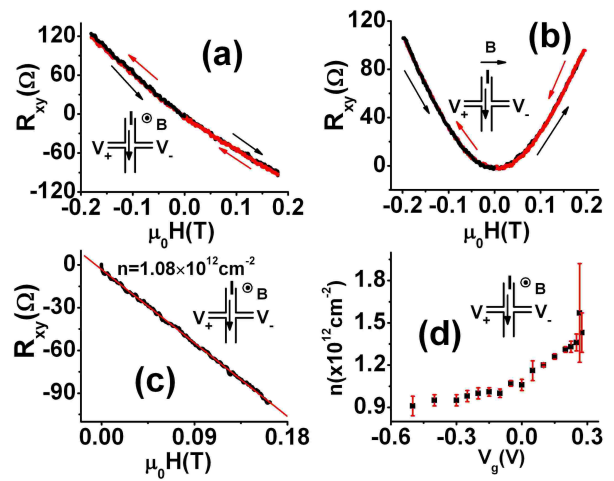}
\centering \caption{No sign of hysteresis is present at T$\sim$4.2K in both (a) perpendicular magnetic field and (b) planar magnetic field (Here the Hall pattern was defined only by chemical etching). The arrows indicate the directionality of the field change. (c) Carrier density of the 2DEG at T$\sim$4.2K. (d) Sheet carrier's density vs gate voltage at 4.2K. Note the change in close near $\sim$0 V, indicating filling of the GaMnAs QW (see Section 4).}

\label{fig:GHB23}

\end{figure}
To further confirm that the carrier type was the electron, we evaporated a metal gate of pure Au, 30 nm thick, and measured the dependence of the carrier's density as a function of the applied voltage (figure 4). It is clearly visible that as we apply a negative voltage, the density decreases while it increases as the voltage is raised to positive values.
\begin{figure}[ht]

\centering \includegraphics[clip, width=3.1in]{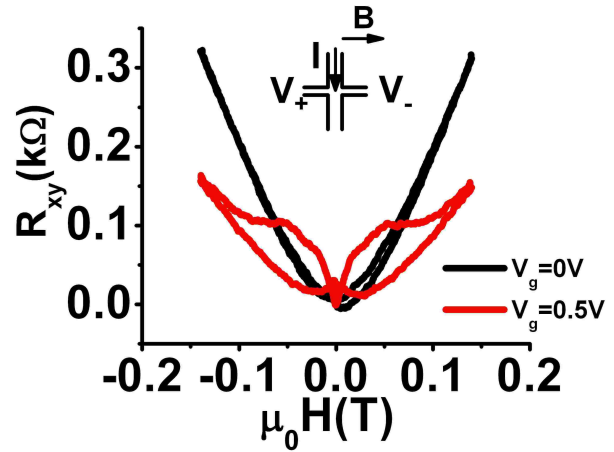}
\centering \caption{GPHE (Hysteretic behaviour at T=0.3K. The direction of the applied magnetic field with respect to the excitation current's direction is indicated in the inset.}

\label{fig:GHB2Hysteresys}

\end{figure}
This is consistent with an n-type doping of the sample. Note also the change in slope near $V_{g}\sim$0 V, indicating the filling of the GaMnAs QW, (See section 4 for discussion).

Subsequently, the sample was cooled below the Curie temperature ($\sim$1.7 K), and an in-plane magnetic field was applied. Initially, no hysteretic behaviour was observable in the Hall signal (black traces in figure 4). After applying a positive gate voltage of 500 mV to fill the GaMnAs QW, a clear hysteretic loop appeared in the hall signal (red traces in figure 4).

One possible interpretation for the presence of the hysteretic loops in the hall signal could be the following. When we first cool down the sample to below the Curie temperature, all the electrons are sitting at the interface between the 200 nm GaAs layer and the 100 nm AlGaAs layer. This means that they are sitting $\sim$110 nm away from where the magnetic impurities are and, therefore, cannot interact. When we apply a positive gate voltage, we manage to fill the quantum well (QW), and at that point, the electrons' wave functions can overlap with the magnetic impurities and the magnetic interaction can actually happen. This interpretation seems to be confirmed by the analysis we present in the following section.

\section{Data Interpretation}
It is essential to convincingly demonstrate that hysteretic behaviour is observable only when the GaMnAs QW is filled with electrons.  Figure 3(d) shows the change of the electron density as a function of the applied gate voltages. If we plot the same data separately, one for the positive gate voltage and one for the negative ones (see figure 5(a-b)), we can see that they present two different slopes which give two different depths for the 2DEG. When we apply positive gate voltages, the differential capacitance indicates that the 2DEG sits at a depth of $\sim52\pm11$ nm (figure 5(b)) which would be where the 10 nm undoped GaMnAs layer is, whereas for negative gate voltages the 2DEG appears to be at a depth of $\sim210\pm54$ nm (figure 5(a)), corresponding more closely to the interface between the 200 nm GaAs and 100 nm AlGaAs layers.
\begin{figure}[ht]

\centering \includegraphics[clip, width=3.1in]{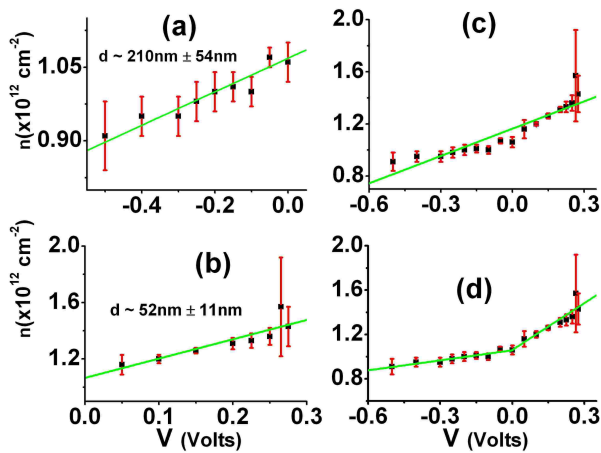}
\centering \caption{GPHE (Sheet carrier's density vs the applied gate voltage. The green lines represent the best fit which gives two different depths for the 2DEG: (a) for negative gate voltages $d\sim210\pm54$ nm and (b) for positive gate voltages $d\sim52\pm11$ nm. One (c) versus two (d) line segment fits.}

\label{fig:GHB2Capacitance}

\end{figure}

To distinguish the scenarios of two versus one slope through all data points, we rely on the statistical F-test, which is useful for comparing two expressions belonging to the same family of expressions but one with more fitting parameters than the other.  We fit to a curve containing one line segment (figure 5(c)), and to a curve containing two line segments (figure 5(d)), and we compare the  -squares of the two fits, using the F random variable \cite{Ftest}:
\begin{equation}
F(n,m)=\frac{(\chi^{2}_{2slopes}-\chi^{2}_{1slope})/\chi^{2}_{2slopes}}{(df_{2slopes}-df_{1slope})/df_{2slopes}}
\end{equation}

Here $n=df_{2slopes}-df_{1slope}$, $m=df_{2slopes}$ with $df$ representing the number of degree of freedom defined by $df=N-p$ , where \textit{N} is the number of data points and \textit{p} the number of fitting parameters. To accept the two line-segment while refusing the one line-segment model with a confidence level exceeding 99$\%$, F(n,m) must exceed the criterion for an integrated probability of 99$\%$ in the F(n,m) distribution with n and m degrees of freedom. Our F-test indicates that the two line-segment model is preferred with a confidence level $\geq$ 99.9$\%$.

To further confirm that indeed we move electrons from the heterojunction to the QW, we simulated our structure with a program that employs the method of finite differences to produce a one-dimensional band diagram together with the hole/electron density \cite{Snider}. 

Figure 6(a) shows the simulated result of the un-gated heterostructure. It can be seen that a triangular well forms at the heterojunction with the presence of occupied electron states below the Fermi level.
\begin{figure}[ht]

\centering \includegraphics[clip, width=3.1in]{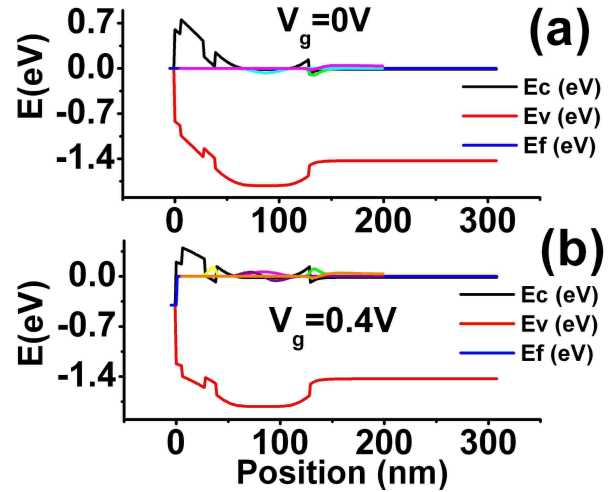}
\centering \caption{GPHE ((a) Unbiased simulated band diagram of our structure. (b) Simulated band diagram with a 400 mV bias applied.}

\label{fig:GHB2Smallpaper2}

\end{figure}

With an applied gate voltage of 0.4 V, we see that the bottom of the QW sat below the Fermi level and that quantized states are immediately occupied, see figure 6(b). All these observations seem to confirm our original hypothesis: when the electrons get close enough to the Mn layer and their wave functions overlap with the magnetic impurities' ones, a ferromagnetic interaction mediated by the electrons present in the QW is realized.

To conclude we can state that we have successfully developed a novel technique that allows us to make ohmic contact to a buried 2D gas in a DMS without destroying the ferromagnetic properties of the sample. This technique has been proven to have a very high yield and work also when higher temperatures are needed to make Ohmic contact. We have also shown that the crystal has two parallel channels for conduction: the heterojunction sitting $\sim$140 nm below the sample's surface and the QW sitting $\sim$40 nm below the sample's surface. The presence of these two conduction channels affects our ability to precisely determine the electron's density thus introducing a significant uncertainty on our density's value. However, the dependence of the carrier's density on the applied gate voltage allows us to estimate this error. Since the capacitance measurement, in figure 2, shows that the 2DEG sits at a depth of about $52\pm11$ nm, closely matching the position of the QW ($\sim$40 nm), we can argue that the density's value measured at zero gate voltage is within 50$\%$, at most, of the real value. 

\section{Acknowledgments}
Angelo Bove would like to thank Ndeye Khady Bove for her support. This work was supported by NSF DMR-0135931.

\end{document}